\RequirePackage[mathlines]{lineno} 
\documentclass[a4paper,11pt]{article}
\pdfoutput=1 

\usepackage{jheppub} 

\usepackage[T1]{fontenc} 

\usepackage{threeparttable}
\usepackage{multirow}
\usepackage{subfigure}
\usepackage{float}
\let\oldequation\equation
\let\oldendequation\endequation
\renewenvironment{equation}
  {\linenomathNonumbers\oldequation}
  {\oldendequation\endlinenomath}

\def \PP  {\emph{PP}}
\def \ee   {e^+e^-}

\def \dsstp  {D_s^{*+}}
\def \dsstm  {D_s^{*-}}
\def \dsp  {D_s^+}
\def \dsm  {D_s^-}

\def \dsstpdsm  {D_s^{*+}D_s^{-}}

\def \kaon {K}
\def \kp  {K^+}
\def \km  {K^-}

\def \piz  {\pi^0}
\def \pip  {\pi^+}
\def \pim  {\pi^-}

\def \etaprim  {\eta'}
\def \ks    {K_{S}^{0}}
\def \dstoketaprim  {D_{s}^+\to\kaon^{+}\eta'}
\def \dstopietaprim  {D_{s}^+\to\eta'\pi^{+}}
\def \dstoketa  {D_s^{+}\to\kaon^{+}\eta}
\def \dstopieta  {D_s^{+}\to\eta\pi^{+}}
\def \dstokpipi  {D_s^{+}\to\kaon^{+}\ks}
\def \dstopipipi  {D_s^{+}\to\ks\pi^{+}}
\def \dstokpiz  {D_s^{+}\to\kaon^{+}\piz}
\def \dstokkpi  {D_s^{+}\to\kp\km\pi^{+}}

\def \dstopp  {D_{s}^{+}\to\PP^{\,\prime}}

\def \pietaprim  {\eta'\pi^{+}}
\def \keta  {\kaon^{+}\eta}
\def \pieta  {\eta\pi^{+}}
\def \kpipi  {\kaon^{+}\ks}
\def \pipipi  {\ks\pi^{+}}
\def \kpiz  {\kaon^{+}\piz}

\def \dedx {\mbox{d}E/\mbox{d}x}
\def \costht {\cos\theta}

\def \deltaE {\Delta \mathbb{E}}
\def \mdspgam {M(\dsp\gamma)}
\def \mdsprec {{M_{\rm rec}}(\dsp)}

\def \mdsp {M(\dsp)}

\def \gev  {\mbox{GeV}}

\def \gevcc{\mbox{GeV/$c^2$}}
\def \mev  {\mbox{MeV}}

\begin{document}

\title{\boldmath Precise Measurements of Branching Fractions for $\dsp$ Meson Decays to Two Pseudoscalar Mesons}
\author{\small
M.~Ablikim$^{1}$, M.~N.~Achasov$^{10,d}$, P.~Adlarson$^{64}$, S. ~Ahmed$^{15}$, M.~Albrecht$^{4}$, A.~Amoroso$^{63A,63C}$, Q.~An$^{60,48}$, ~Anita$^{21}$, Y.~Bai$^{47}$, O.~Bakina$^{29}$
% ...
\vspace{0.2cm}\\
(BESIII Collaboration)\\
\vspace{0.2cm} {\it
$^{1}$ Institute of High Energy Physics, Beijing 100049, People's Republic of China\\
% ...
$^{4}$ Bochum Ruhr-University, D-44780 Bochum, Germany\\
% ...
$^{10}$ G.I. Budker Institute of Nuclear Physics SB RAS (BINP), Novosibirsk 630090, Russia\\
% ...
$^{15}$ Helmholtz Institute Mainz, Johann-Joachim-Becher-Weg 45, D-55099 Mainz, Germany\\
% ...
$^{21}$ Indian Institute of Technology Madras, Chennai 600036, India\\
% ...
$^{29}$ Joint Institute for Nuclear Research, 141980 Dubna, Moscow region, Russia\\
% ...
$^{47}$ Southeast University, Nanjing 211100, People's Republic of China\\
$^{48}$ State Key Laboratory of Particle Detection and Electronics, Beijing 100049, Hefei 230026, People's Republic of China\\
% ...
$^{60}$ University of Science and Technology of China, Hefei 230026, People's Republic of China\\
% ...
$^{63}$ (A)University of Turin, I-10125, Turin, Italy; (B)University of Eastern Piedmont, I-15121, Alessandria, Italy; (C)INFN, I-10125, Turin, Italy\\
$^{64}$ Uppsala University, Box 516, SE-75120 Uppsala, Sweden\\
% ...
$^{d}$ Also at the Novosibirsk State University, Novosibirsk, 630090, Russia\\
% ...
}\vspace{0.4cm}}

\abstract{
We measure the branching fractions for seven $D_{s}^{+}$ two-body decays to pseudo-scalar mesons, by analyzing data collected at $\sqrt{s}=4.178\sim4.226~\gev$ with the BESIII detector at the BEPCII collider. The branching fractions are determined to be
\begin{center}
	$\mathcal{B}(D_s^+\to\kaon^+\eta^{\prime})=(2.68\pm0.17\pm0.17\pm0.08)\times10^{-3}$,\\
	$\mathcal{B}(D_s^+\to\eta^{\prime}\pi^+)=(37.8\pm0.4\pm2.1\pm1.2)\times10^{-3}$,\\
	$\mathcal{B}(D_s^+\to\kaon^+\eta)=(1.62\pm0.10\pm0.03\pm0.05)\times10^{-3}$,\\
	$\mathcal{B}(D_s^+\to\eta\pi^+)=(17.41\pm0.18\pm0.27\pm0.54)\times10^{-3}$, \\
	$\mathcal{B}(D_s^+\to\kaon^+\kaon_S^0)=(15.02\pm0.10\pm0.27\pm0.47)\times10^{-3}$,\\
	$\mathcal{B}(D_s^+\to\kaon_S^0\pi^+)=(1.109\pm0.034\pm0.023\pm0.035)\times10^{-3}$, \\
	$\mathcal{B}(D_s^+\to\kaon^+\pi^0)=(0.748\pm0.049\pm0.018\pm0.023)\times10^{-3}$,
\end{center}
where the first uncertainties are statistical, the second are systematic, and the third are from external input branching fraction of the normalization mode $\dstokkpi$. Precision of our measurements is significantly improved compared with that of the current world average values.
}

\maketitle
\flushbottom

\section{INTRODUCTION}
\label{sec:intro}

Among the hadronic decays of the strange-charmed meson $D_s^+$, the theoretical treatment based on QCD-inspired models of its decays into two pseudoscalar mesons ($\dstopp$) is the cleanest~\cite{Hai-Yang Cheng2010,Fu-Sheng Yu2011}.
Precision measurements of these decay rates can provide crucial calibrations to different theoretical models~\cite{Cheng:2019ggx,Hai-Yang Cheng2010,Fu-Sheng Yu2011,Hsiang-nan Li2012,Di Wang2017}.
For each decay branching fraction (BF) listed in Table~\ref{tab:results_BFs_theo}, the precision of current measurements listed by the Particle Data Group (PDG)~\cite{pdg2018} is still not good enough to test theoretical models. Hence, more precise and independent measurements are desired to further improve our understanding of QCD dynamics in charm physics.

In 2019, LHCb discovered $\emph{CP}$ violation in $D^0 \to \pi^+ \pi^-$ and $D^0\to K^+K^-$ decays with a significance of 5.3$\sigma$~\cite{Aaij:2019kcg}, providing stringent constraints on theoretical approaches to $\emph{CP}$ violation in the charm sector~\cite{Hai-Yang Cheng2010,Hsiang-nan Li2012,Miroslav:2020sb}.  
For the strange-charmed meson $D_s^+$, there are theoretical predictions for the $\emph{CP}$ asymmetries of the singly Cabibbo-suppressed (SCS) decay modes, which rely on the potential effect of SU(3) symmetry breaking~\cite{Cheng:2019ggx, Buccella:2019kpn}.
However, the current world average results, as shown in Table~\ref{tab:results_BFs_theo}, suffer from large uncertainties and are thus insensitive to SU(3) breaking. 
More precise measurements of the BFs for the SCS modes in $\dstopp$ will help to explore SU(3) symmetry breaking in $D_s^+$ decays~\cite{Cheng:2019ggx, Buccella:2019kpn}. As a result, more reliable theoretical predictions of $\emph{CP}$ asymmetries in the $D_s^+$ SCS hadronic decays can be achieved. 

In this work, we measure the BFs for seven two-body hadronic decays $\dstopp$: $\dstoketaprim$, $\pietaprim$, $\keta$, $\pieta$, $\kpipi$, $\pipipi$ and $\kpiz$. These decay modes were previously measured by CLEO~\cite{cleo2008, cleo2010, cleo2013}. The analysis is carried out in the process of $e^+e^-\to D_s^+D_s^{*-} + c.c.\to \gamma D_s^+ D_s^-$ based on data samples collected at the center-of-mass energies $\sqrt{s}$ = 4.178, 4.189, 4.199, 4.209, 4.219 and 4.226~$\gev$, corresponding to the integrated luminosities of 3189.0, 526.7, 526.0, 517.1, 514.6 and 1091.7~pb$^{-1}$, respectively~\cite{Ablikim:2015zaa, Ablikim:2015nan}.

A partial reconstruction technique is adopted: only one $D_s^{\pm}$, decaying into the  $PP^{\,\prime}$
mode, is detected along with a soft photon from $D_s^{*\pm}(D_s^{*\mp})$; 
the other $D_s^{\mp}$ is not used.  The BFs are measured relative to the normalization mode $D_s^+\to K^+K^-\pi^+$. In the context, charge conjugate modes are always implied, unless explicitly mentioned.

\begin{table*}[h]
	\setlength{\abovecaptionskip}{0.cm}
	\setlength{\belowcaptionskip}{-0.2cm}
	\caption{Comparisons of the $D_s^+$ decay BFs between the world average results from PDG~\cite{pdg2018} and calculations from different theoretical models (in unit of $10^{-3}$).}
	\label{tab:results_BFs_theo}
	\newcommand{\tabincell}[2]{\begin{tabular}{@{}#1@{}}#2\end{tabular}} %
	\begin{center}
		\footnotesize
		\begin{threeparttable}
			\begin{tabular}{c|c c c c c c c}
				\hline\hline
				\multirow{2}{*}{Decay} & \multirow{2}{*}{\tabincell{c}{PDG\\~\cite{pdg2018}}} &\multicolumn{2}{c}{ \tabincell{c}{Cheng $\emph{et al.}$~\cite{Cheng:2019ggx}}}& \multirow{2}{*}{\tabincell{c}{Cheng $\emph{et al.}$\\~\cite{Hai-Yang Cheng2010}}} & \multirow{2}{*}{\tabincell{c}{Yu $\emph{et al.}$\\~\cite{Fu-Sheng Yu2011}}}  & \multirow{2}{*}{\tabincell{c}{Li $\emph{et al.}$\\~\cite{Hsiang-nan Li2012}}} & \multirow{2}{*}{\tabincell{c}{Wang $\emph{et al.}$\\~\cite{Di Wang2017}}} \\
				&&SU(3)&\tabincell{c}{SU(3)-breaking}&&&&\\
				\hline
				$\kaon^+\eta'$     & $1.8\pm0.6$   & \tabincell{c}{$1.23\pm0.06$} & \tabincell{c}{$1.49\pm0.08$} & \tabincell{c}{$1.07\pm0.17$}   & \tabincell{c}{$1.4\pm0.4$} & $1.92$ & \tabincell{c}{$3.1\pm0.4$} \\
				$\eta'\pi^+$       & $39.4\pm2.5$   & -                                  & -                                  & \tabincell{c}{$38.2\pm3.6$}     & \tabincell{c}{$46\pm6$} & $34.4$ & \tabincell{c}{$46.7\pm6.2$}\\
				$\kaon^+\eta$      & $1.77\pm0.35$ & \tabincell{c}{$0.91\pm0.03$} & \tabincell{c}{$0.86\pm0.03$} & \tabincell{c}{$0.78\pm0.09$}   & \tabincell{c}{$0.8\pm0.5$} & $1.00$ & \tabincell{c}{$0.91\pm0.20$} \\
				$\eta\pi^+$        & $17.0\pm0.9$   & -                                  & -                                  & \tabincell{c}{$18.2\pm3.2$}     & \tabincell{c}{$19\pm5$} & $16.5$ & \tabincell{c}{$19.6\pm4.4$} \\
				$\kaon^+\ks$       & $15.0\pm0.5$   & -                                  & -                                  & \tabincell{c}{$14.85\pm1.60$}   & \tabincell{c}{$15.0\pm4.5$} & $15.0$ & \tabincell{c}{$15.0\pm1.6$} \\
				$\ks\pi^+$         & $1.22\pm0.06$ & \tabincell{c}{$1.20\pm0.04$} & \tabincell{c}{$1.27\pm0.04$} & \tabincell{c}{$1.365\pm0.130$} & \tabincell{c}{$1.4\pm0.3$} & $1.105$ & \tabincell{c}{$1.04\pm0.13$} \\
				$\kaon^+\pi^{0}$   & $0.63\pm0.21$ & \tabincell{c}{$0.86\pm0.04$} & \tabincell{c}{$0.56\pm0.02$} & \tabincell{c}{$0.86\pm0.09$}   & \tabincell{c}{$0.5\pm0.2$} & $0.67$ & \tabincell{c}{$0.69\pm0.03$} \\
				\hline\hline
			\end{tabular}
		\end{threeparttable}
	\end{center}
\end{table*}

\section{BESIII DETECTOR AND MONTE CARLO SIMULATION}
The BESIII detector is a magnetic spectrometer~\cite{Ablikim:2009aa} located at BEPCII~\cite{Yu:IPAC2016-TUYA01}. The cylindrical core of the BESIII detector consists of a helium-based main drift chamber (MDC), a plastic scintillator time-of-flight system (TOF), and a CsI(Tl) electromagnetic calorimeter (EMC), which are all enclosed in a superconducting solenoidal magnet providing a 1.0~T magnetic field. The solenoid is supported by an octagonal flux-return yoke with resistive plate counter muon tracker modules interleaved with steel. The acceptance of
charged particles and photons is 93\% over $4\pi$ solid angle. The charged-particle momentum resolution at $1~{\rm GeV}/c$ is $0.5\%$, and the ionization energy loss~$\dedx$ resolution is $6\%$ for the electrons from Bhabha scattering. The EMC measures photon energies with a resolution of $2.5\%$ ($5\%$) at $1$~GeV in the barrel (end cap) region. The time resolution of the TOF barrel part is 68~ps, while that of the end cap part is 110~ps. The end cap TOF system was upgraded in 2015 with multi-gap resistive plate chamber technology, providing a time resolution of 60~ps~\cite{etof,etof1}. Only the 4.226 GeV data was taken before this upgrade.  

Simulated data samples, produced with the {\sc geant4}-based~\cite{geant4} Monte Carlo (MC) package which includes the geometric description of the BESIII detector and the detector response, are used to determine the detection efficiency and to estimate the backgrounds. The simulation includes the beam energy spread and initial state radiation in the $e^+e^-$ annihilations modeled with the generator {\sc kkmc}~\cite{ref:kkmc, ref:kkmc1}. In order to study the backgrounds, generic MC samples consisting of open-charm states, radiative return to $J/\psi$ and $\psi(2S)$, and continuum processes of $q\bar{q}~(q=u, d, s)$, along with Bhabha scattering, $\mu^+\mu^-$, $\tau^+\tau^-$, and $\gamma\gamma$ events are generated.
The known decay modes are modeled with {\sc evtgen}~\cite{ref:evtgen, ref:evtgen1} using BFs taken from PDG~\cite{pdg2018}, and the remaining unknown decays from the charmonium states are treated with {\sc lundcharm}~\cite{ref:lundcharm,ref:lundcharm1}.  Final state radiation (FSR) from charged final state particles is incorporated with the {\sc photos} package~\cite{photos}. The signal MC samples of $\ee\to D_s^{*\pm}D_s^{\mp}$ with a $\dsp$ meson decaying to the signal decay modes together with a $\dsm$ decaying inclusively are generated with {\sc ConExc}~\cite{Ping:2013jka}.

\section{MEASUREMENT METHOD}	
\label{sec:method}
In this analysis, a candidate $\dsp$ meson is reconstructed by the combination of the detected final-state particles. With current precision, $\emph{CP}$ violation is negligible, which means the BFs for $\dsp$ decays to the mode $i^+$, $\mathcal{B}^{i^+}\equiv\mathcal{B}(D_s^+ \to i^+)$, and for  $\dsm$ decays to the mode $i^-$, $\mathcal{B}^{i^-}$, are equal. Therefore, we denote $\mathcal{B}^{i^+}=\mathcal{B}^{i^-}=\mathcal{B}^{i}$. The yield, $n^{i}$, of the observed $\dsp\to i$ signal events at all six energy points can be written as
\begin{equation}
	n^{i}=2N^{D_s^{*+}D_s^{-}} \cdot\mathcal{B}^i\cdot\mathcal{B}_{\rm {final-state}}^{i}\cdot\overline{\varepsilon}^i,
	\label{eq:signal}
\end{equation}
where $N^{D_s^{*+}D_s^{-}}$ is the total number of $D_s^{*+}D_s^{-}$ pairs produced in all the data samples. For mode $i$, $\mathcal{B}_{\rm {final-state}}^i$ is the combined BF from the $i$ state to the observed final state (those are described in Section 4), and $\overline{\varepsilon}^i$ is average detection efficiency for the whole data set, which is given as
\begin{equation}
	\overline{\varepsilon}^i = \frac{ \sum\limits_{k=1}^6  L_{k}\cdot\sigma_{k} \cdot \varepsilon_k^i}{\sum\limits_{k=1}^6L_{k} \cdot \sigma_{k}}.
	\label{eq:eff_weight}
\end{equation}
Here, $L_{k}$ is the integrated luminosity, $\sigma_{k}$ is the observed cross section and $\varepsilon_k^i$ is the detection efficiency at the $k$-th energy point.

The absolute BF of the normalization mode decay, $\dstokkpi$, is denoted by $\mathcal{B}^{\kaon^+\kaon^-\pi^+}$ and is taken from PDG~\cite{pdg2018}.  Based on Eq.~\eqref{eq:signal}, the relative BF for the signal mode $D_s^+\to i$ is  
\begin{equation}
	\begin{aligned}
		\emph{$R^{i}$}=\frac{\mathcal{B}^i}{\mathcal{B}^{K^+K^-\pi^+}}= \frac{n^i \cdot \overline{\varepsilon}^{K^+K^-\pi^+}}{n^{K^+K^-\pi^+} \cdot \overline{\varepsilon}^i \cdot \mathcal{B}^i_{\rm {final-state}}}\,.
		\label{eq:relative_ratio}
	\end{aligned}
\end{equation}
The absolute BF $\mathcal{B}^{i}$ is obtained by
\begin{equation}
	\begin{aligned}
		\mathcal{B}^i=R^i\cdot\mathcal{B}^{K^+K^-\pi^+}.
	\end{aligned}
\end{equation}

\section{EVENT SELECTION}
\label{sec:event_selection}
Charged tracks are reconstructed from hits in the MDC. Except for the tracks used to reconstruct the $\ks$ meson, the distances of closest approach to the interaction point are required to satisfy $R_{xy}<1.0$~cm in the $xy$ plane perpendicular to the $z$ direction of the MDC and $R_z<10.0$~cm along the $z$ direction. The track polar angle $\theta$ must satisfy $|\costht|<0.93$. 
For particle identification~(PID) of charged tracks, measurements of $\dedx$ and the flight time measured by the TOF are combined to form a likelihood $L(h)$ ($h=\pi, K$) for each hadron hypothesis. Tracks are identified as charged pions when the PID likelihoods of pions are larger than those of kaons, $L(\pi)>L(K)$, while tracks with $L(K)>L(\pi)$ are identified as kaons.

Shower clusters with no association to any charged tracks in the EMC crystals will be identified as photon candidates when the following requirements are fulfilled: the measured EMC time is within $0\leqslant t \leqslant 700$ ns of the event start time to suppress the electronic noise and showers unrelated to the events; the deposited energy is larger than 25~$\mev$ in the barrel~($|\costht|<0.80$) and larger than 50~$\mev$ in the end cap~($0.86<|\costht|<0.92$). Additionally, the angle between a photon candidate and the nearest charged track must be larger than 10$^\circ$ to prevent contamination from hadronic showers.  

The $\piz$ and $\eta$ meson candidates are reconstructed from photon pairs with the invariant mass $M(\gamma\gamma)$ within [0.120, 0.145]~$\gevcc$ and [0.510, 0.560]~$\gevcc$, respectively. In order to improve the momentum resolution, a kinematic fit constraining the reconstructed $\piz$~($\eta$) mass to its nominal mass~\cite{pdg2018} is applied and the fitted four-momentum of the $\piz$~($\eta$) meson is used for further analysis. The $\etaprim$ meson candidates are reconstructed from $\pip\pim\eta$ with an $M(\pip\pim\eta)$ invariant mass requirement of [0.945, 0.970]~$\gevcc$.

Candidate $\ks$ mesons are reconstructed from two oppositely charged tracks, with no PID requirement; these tracks are required to satisfy the polar angle requirement $|\cos\theta|<0.93$ and $R_{xy}<20$~cm. Furthermore, there is usually a detectable displacement before the decay of $\ks$ meson due to its relatively long lifetime. Therefore, the decay length and corresponding uncertainty of $\ks$ candidates are required to satisfy $L/\sigma_{L} > 2$, which suppresses prompt $\pip\pim$ combinatorial background~\cite{Ks0_rec}. The $\ks$ meson candidates with an invariant mass $M(\pip\pim)$ within the mass window [0.491, 0.505]~$\gevcc$ are retained. 

For a specific $\dsp$ decay mode, the $\dsp$ signal candidates are formed by combining all the detected final-state particles.  In addition, a radiative photon from the $D_s^{*\pm}$ decay must be detected. 
Among all the  $\gamma\dsp$  combinations in the event, the one with the minimal $|\deltaE|$ is kept for subsequent analysis only, where $\deltaE$ is the difference between the center-of-mass energy $E_{0}\equiv \sqrt{s}$ and the total energy of $\gamma\dsp\dsm$ in the center-of-mass frame of the $e^+e^-$ beams
\begin{equation}
	\begin{aligned}
		\deltaE =  (E_{\dsp} + E_{\gamma}+E_{\rm rec}) - E_{0}.
		\label{eq:deltaE}
	\end{aligned}
\end{equation}
Here $E_{\dsp}$ and $E_{\gamma}$ are the energies of reconstructed $\dsp$ and $\gamma$ from $D_s^{*\pm}$, respectively. $E_{\rm rec}$ is the energy of the recoiled $\dsm$, calculated using
\begin{equation}
	\begin{aligned}
		E_{\rm rec}=\sqrt{\left|-(\overrightarrow{p}_{\dsp}+\overrightarrow{p}_{\gamma})\right|^2+m_{\dsm}^2},
		\label{eq:recoilE}
	\end{aligned}
\end{equation}
where \textbf{$\overrightarrow{p}_{\dsp}$} is the total momentum of the detected $\dsp$, \textbf{$\overrightarrow{p}_{\gamma}$} is the momentum of the radiative photon $\gamma$, and $m_{\dsm}$ is the nominal mass of the $\dsm$~\cite{pdg2018}. For a correctly reconstructed $\dsp$ candidate, $\deltaE$ is expected to be around zero. Therefore, candidates will be rejected when they fail the requirements of $\deltaE$ for each decay mode, as shown in Table~\ref{tab:deltaE_cross_cut_ranges}, which correspond to the $\pm3\sigma$ regions of the signal $\deltaE$ distributions. To further improve the kinematic resolutions of the final states, a kinematic fit is performed to constrain the recoil mass of the $\dsp\gamma$, $M_{\rm rec}(\dsp\gamma)$, to the nominal mass of the $\dsm$. According to the kinematic fit, the four momenta of all the final-state particles are updated.

As an example, data for $\dstokpiz$ is shown in Fig.~\ref{fig:crossCut_after_1c_kpi0}; the two-dimensional distribution of the recoil mass $\mdsprec$ and the invariant mass $M(\dsp\gamma)$
depicts the two resonance structures of the processes. The horizontal band corresponds to $\ee\to\dsstp\dsm\to\gamma\dsp\dsm$, while the vertical band corresponds to  $\ee\to\dsp\dsstm\to\dsp\gamma\dsm$. To improve the signal-to-background ratio, we further retain only events lying in the regions of the horizontal or vertical bands defined in Table~\ref{tab:deltaE_cross_cut_ranges}. 

\begin{table}[tp]
	\setlength{\abovecaptionskip}{1.2cm}
	\setlength{\belowcaptionskip}{0.2cm}
	\begin{center}
		\caption{Summary of the requirements of $\deltaE$, $\mdsprec$ and $\mdspgam$ for each $\dstopp$ decay mode and the normalization mode.}
		\vspace{-0.0cm}
		\footnotesize
		\begin{tabular}{c|c|c|c} \hline\hline
			Decay          & $\deltaE$($\gev$) &   $\mdsprec$($\gevcc$) &  $\mdspgam$($\gevcc$) \\ \hline
			$\kaon^+\eta'$   & ($-$0.040, 0.025) & (2.100, 2.130)  &  (2.095, 2.130)\\
			$\eta'\pi^+$     & ($-$0.040, 0.025) & (2.100, 2.130)  &  (2.095, 2.130)\\
			$\kaon^+\eta$    & ($-$0.045, 0.025) & (2.100, 2.130)  &  (2.095, 2.130)\\
			$\eta\pi^+$      & ($-$0.045, 0.025) & (2.100, 2.130)  &  (2.095, 2.130)\\
			$\kaon^+\ks$     & ($-$0.040, 0.020) & (2.100, 2.130)  &  (2.100, 2.130)\\
			$\ks\pi^+$       & ($-$0.040, 0.020) & (2.100, 2.130)  &  (2.100, 2.130)\\
			$\kaon^+\piz$    & ($-$0.050, 0.020) & (2.100, 2.130)  &  (2.100, 2.130)\\
			$\kp\km\pi^+$    & ($-$0.030, 0.020) & (2.100, 2.130)  &  (2.100, 2.130)\\
			\hline\hline
		\end{tabular}
		\label{tab:deltaE_cross_cut_ranges}
	\end{center}
\end{table}
\vspace{-0.0cm}
\begin{figure}[tp] \centering
	\setlength{\abovecaptionskip}{-1pt}
	\setlength{\belowcaptionskip}{10pt}
	\includegraphics[width=10.0cm]{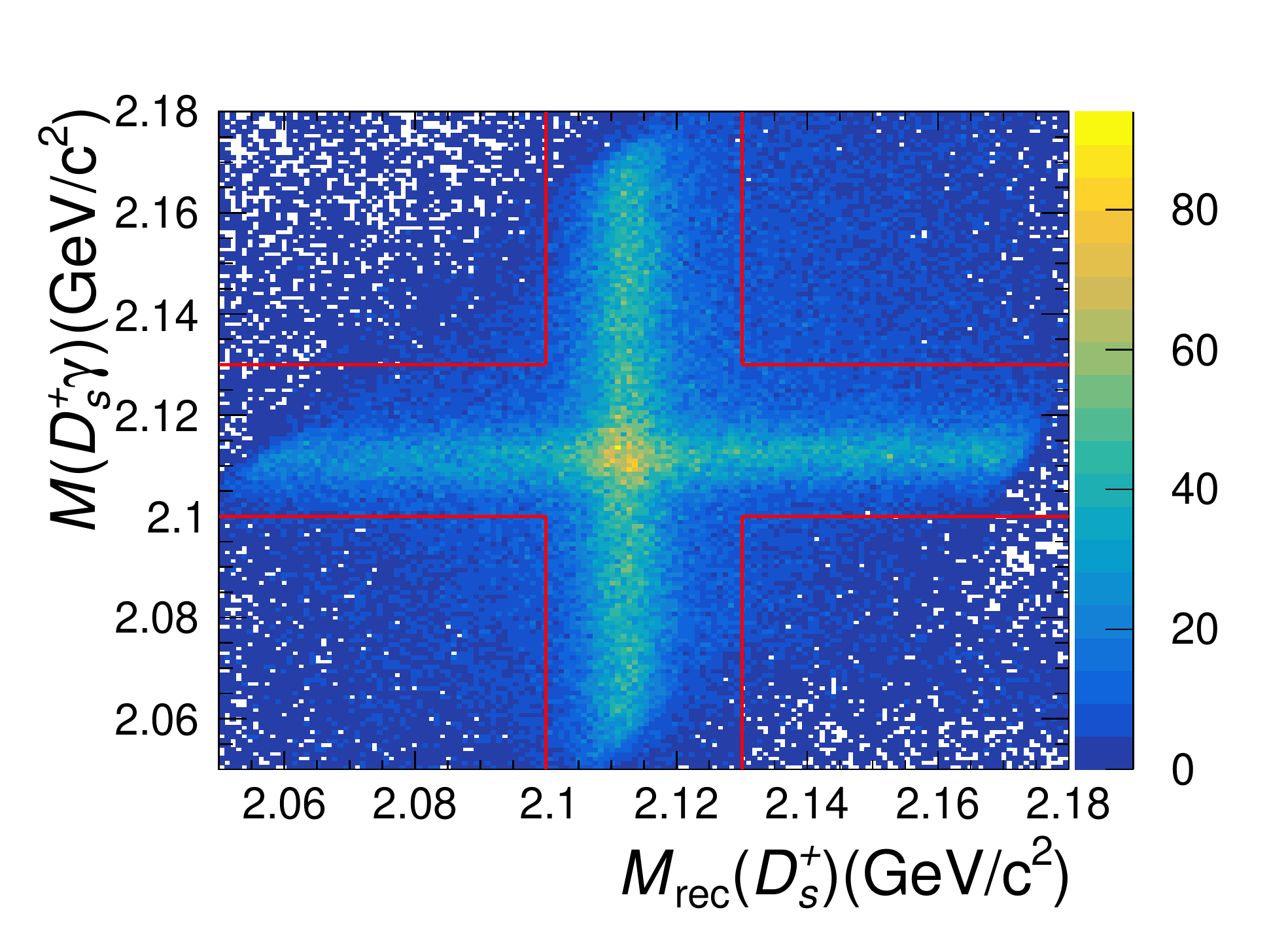}
	\caption{Two-dimensional distribution of the recoil mass of $\dsp$ and the invariant mass of $\dsp\gamma$ for the decay $\dstokpiz$, where the solid lines denote the boundaries for the horizontal and vertical band ranges.}
	\label{fig:crossCut_after_1c_kpi0}
\end{figure}
\vspace{-0.0cm}

\section{SIGNAL YIELD AND BRANCHING FRACTION}
\begin{figure*}[htpb]
	\centering
	\subfigure{
		\begin{minipage}{0.4\columnwidth}
			\centering
			\includegraphics[width=2.5in]{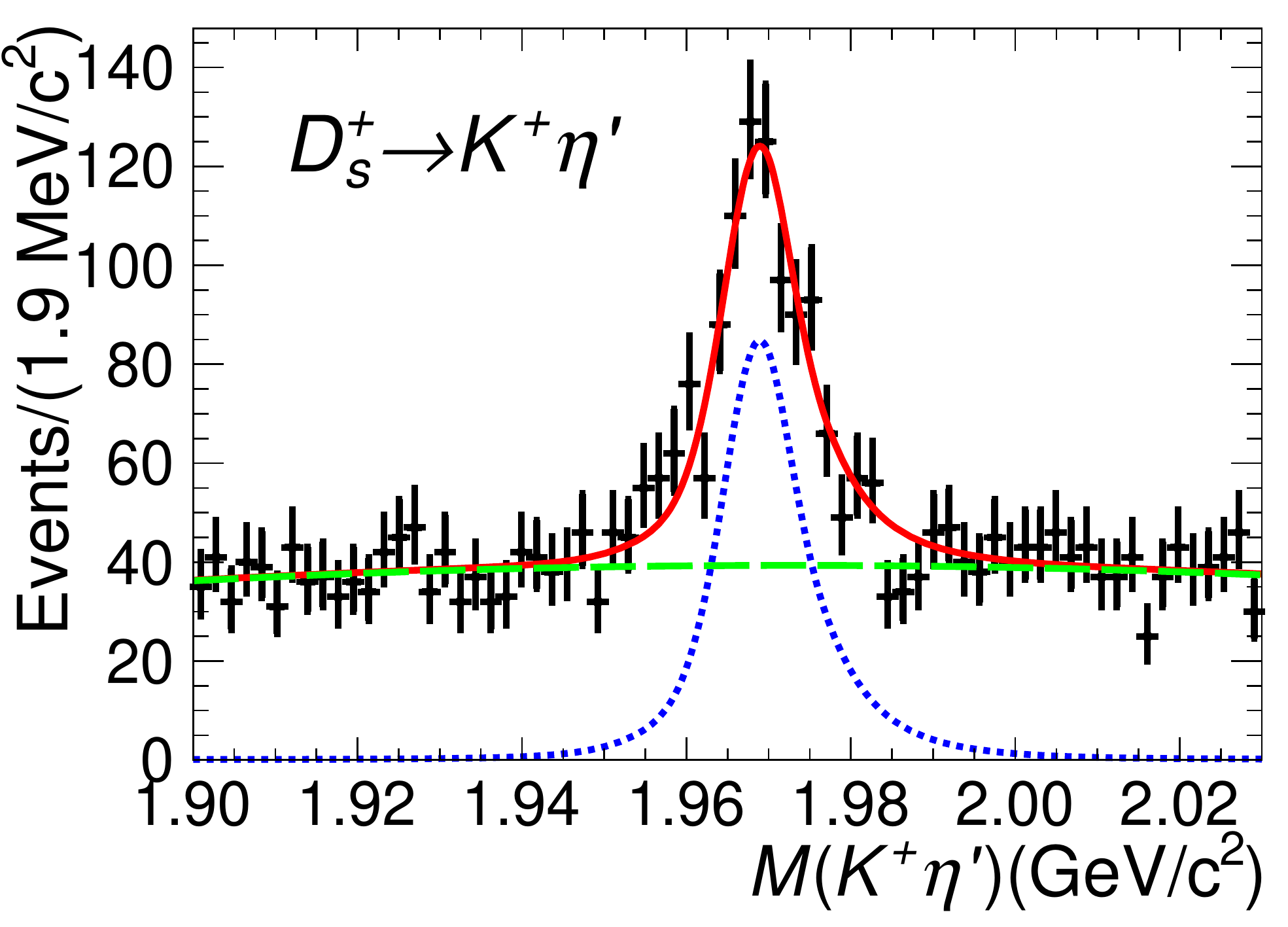}
		\end{minipage}
	}
	\subfigure{
		\begin{minipage}{0.4\columnwidth}
			\centering
			\includegraphics[width=2.5in]{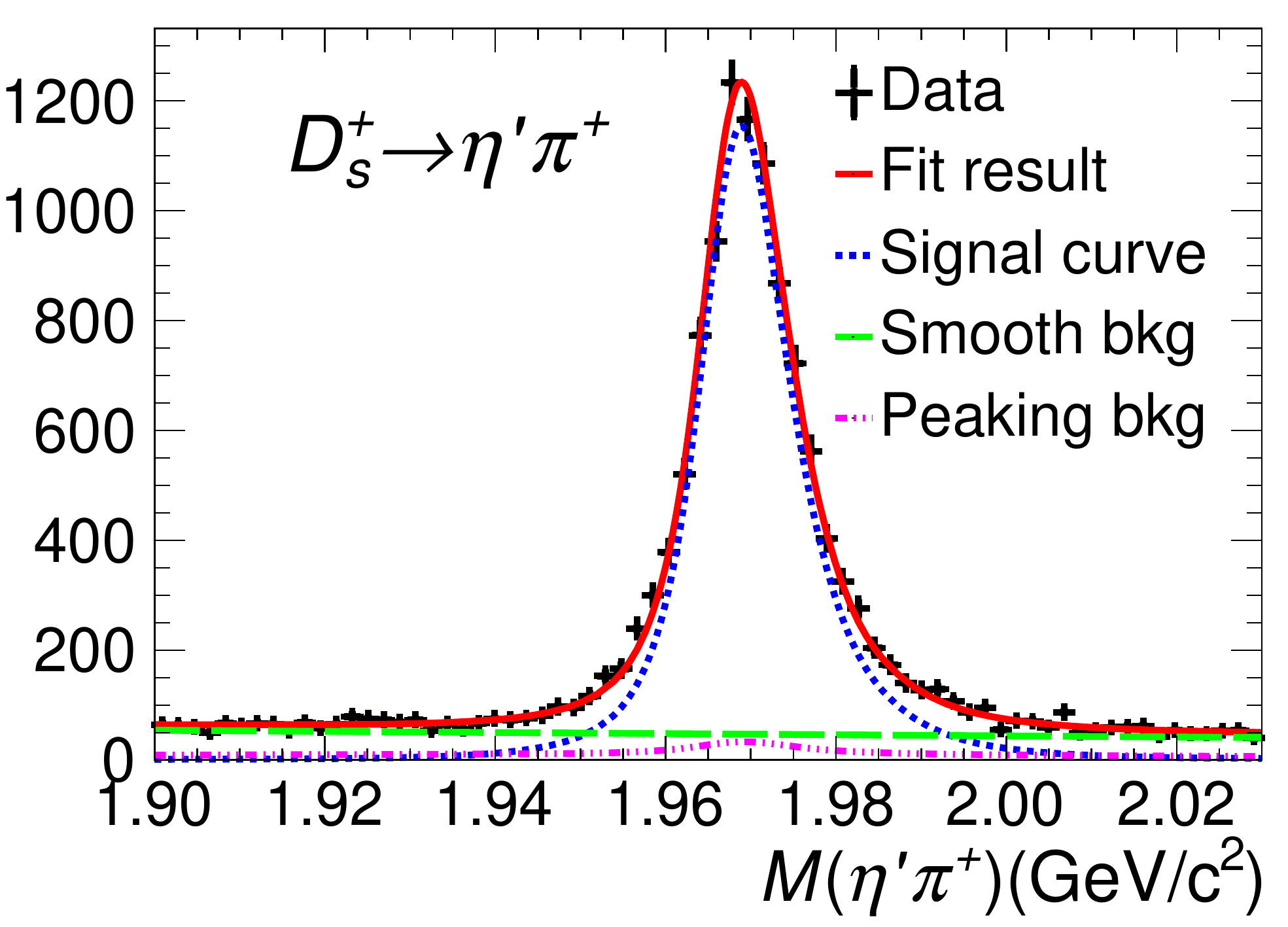}
		\end{minipage}
	}
	\subfigure{
		\begin{minipage}{0.4\columnwidth}
			\centering
			\includegraphics[width=2.5in]{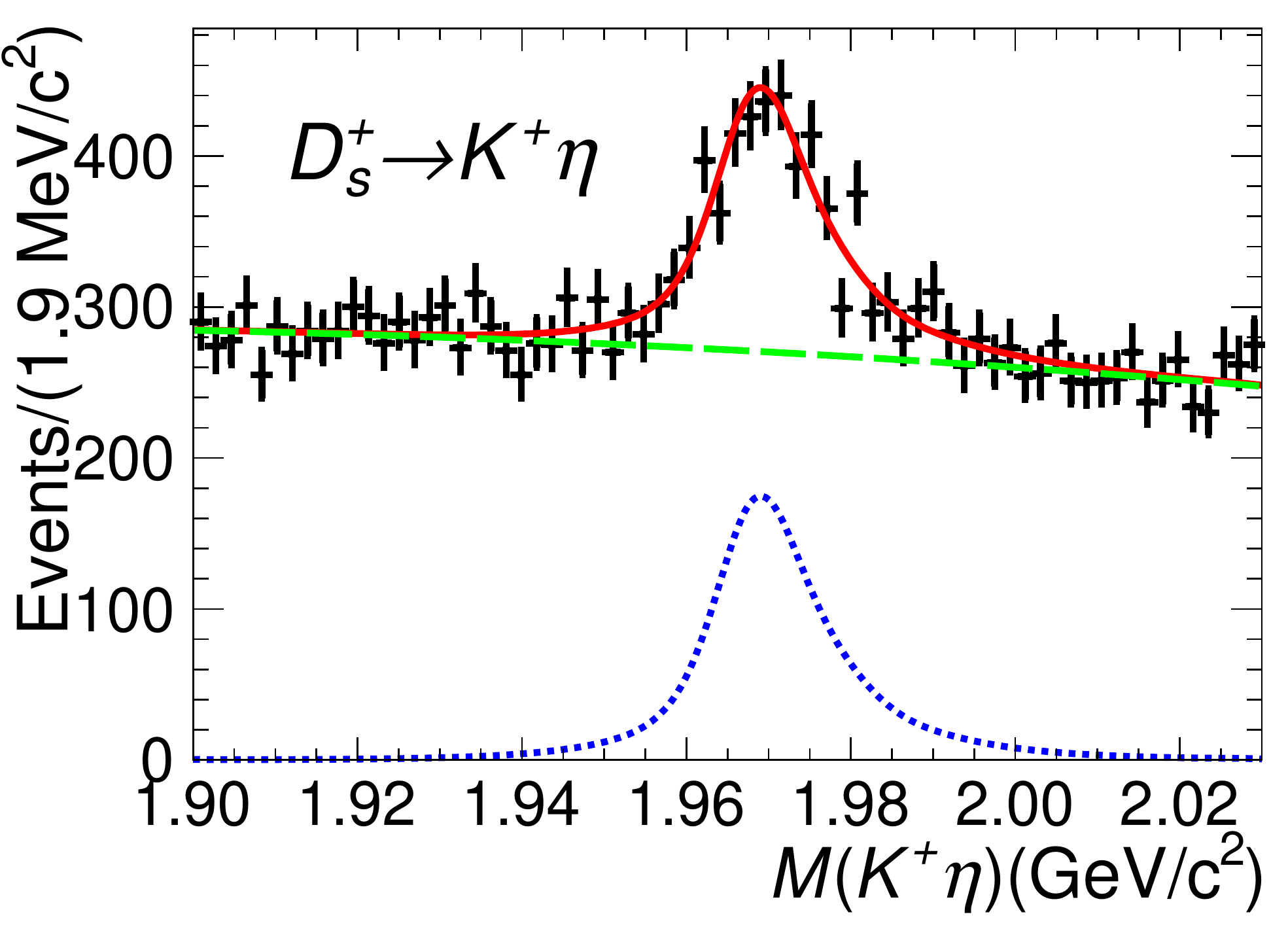}
		\end{minipage}
	}
	\subfigure{
		\begin{minipage}{0.4\columnwidth}
			\centering
			\includegraphics[width=2.5in]{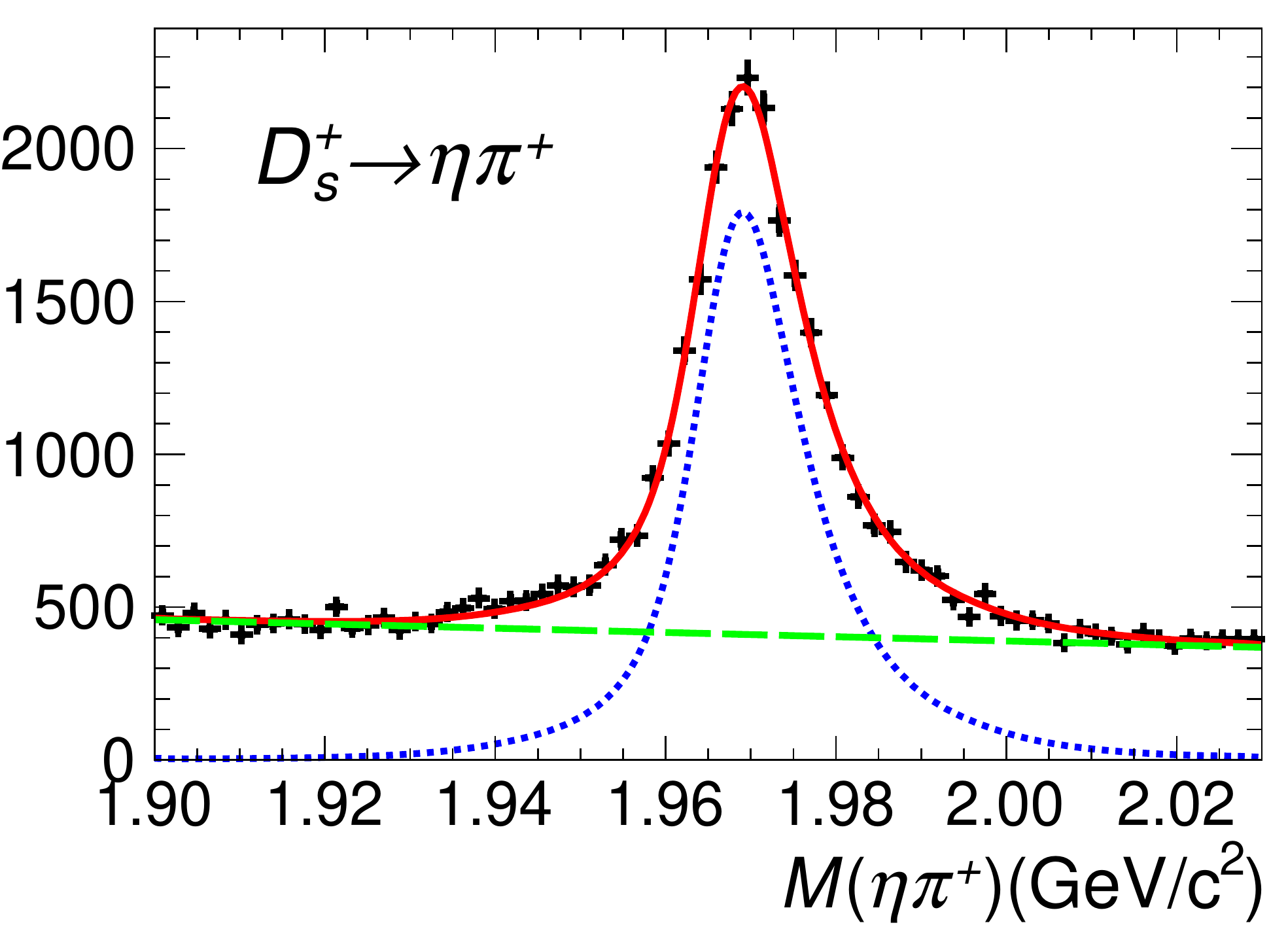}
		\end{minipage}
	}
	\subfigure{
		\begin{minipage}{0.4\columnwidth}
			\centering
			\includegraphics[width=2.5in]{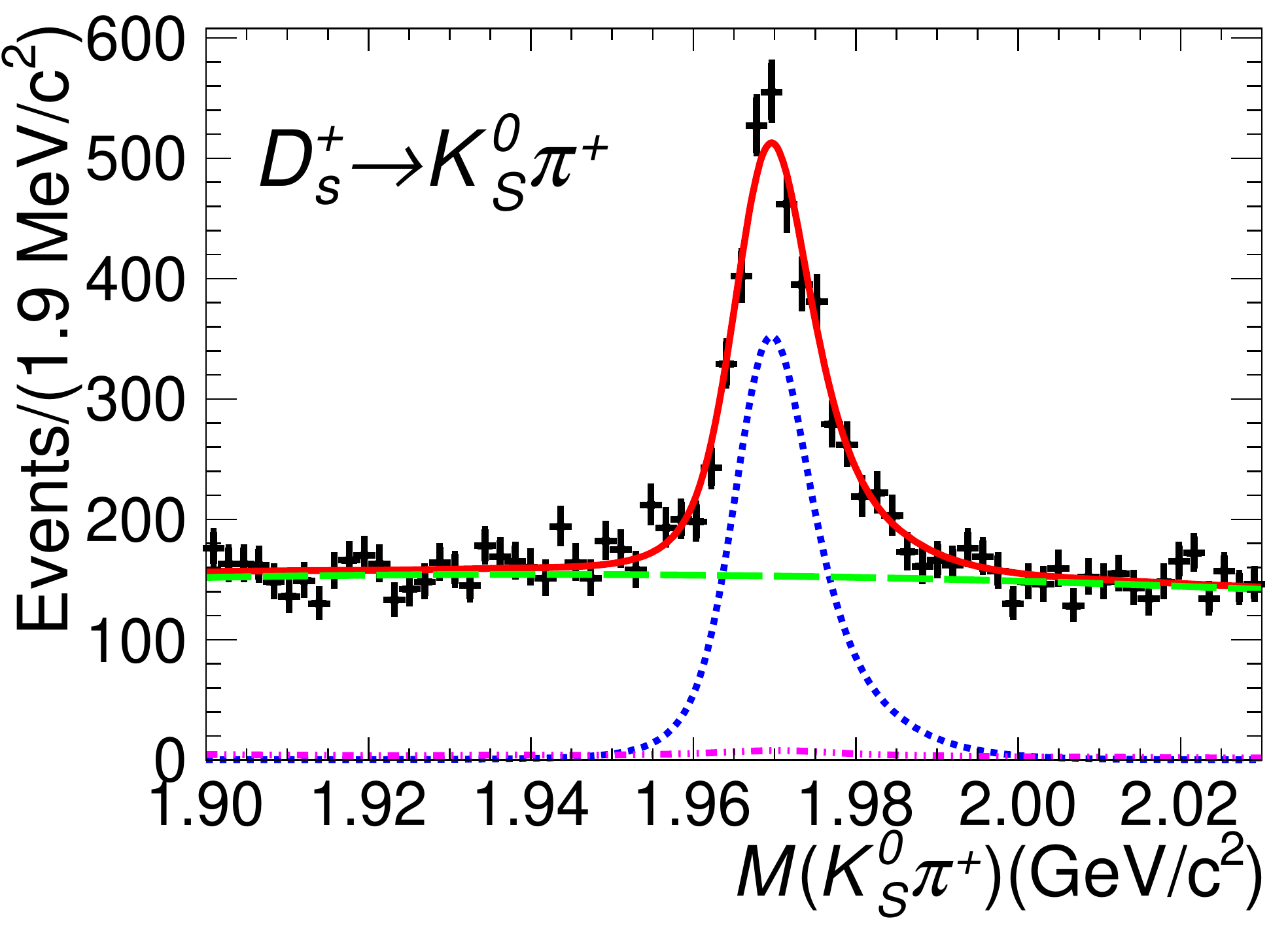}
		\end{minipage}
	}
	\subfigure{
		\begin{minipage}{0.4\columnwidth}
			\centering
			\includegraphics[width=2.5in]{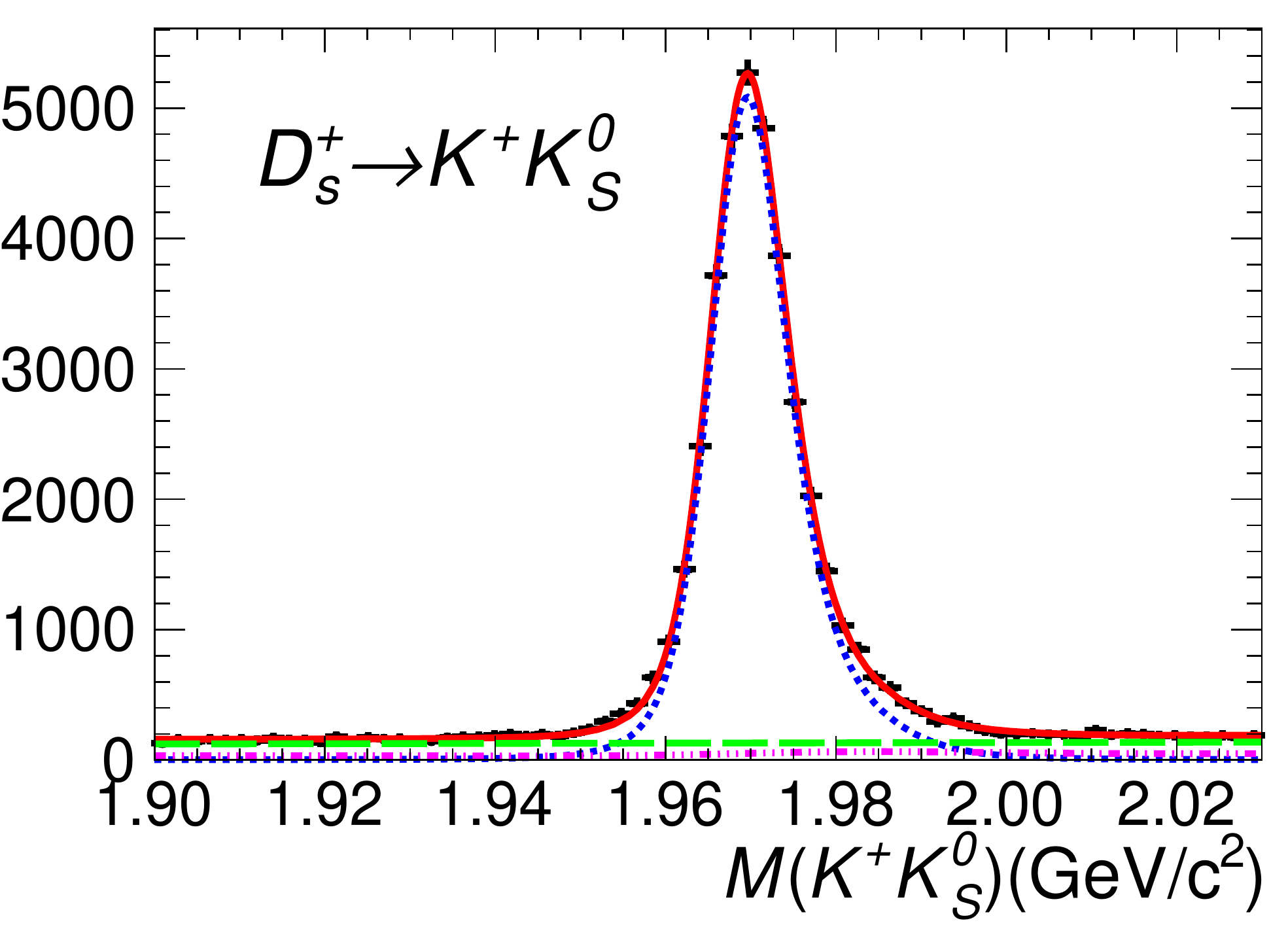}
		\end{minipage}
	}
	\subfigure{
		\begin{minipage}{0.4\columnwidth}
			\centering
			\includegraphics[width=2.5in]{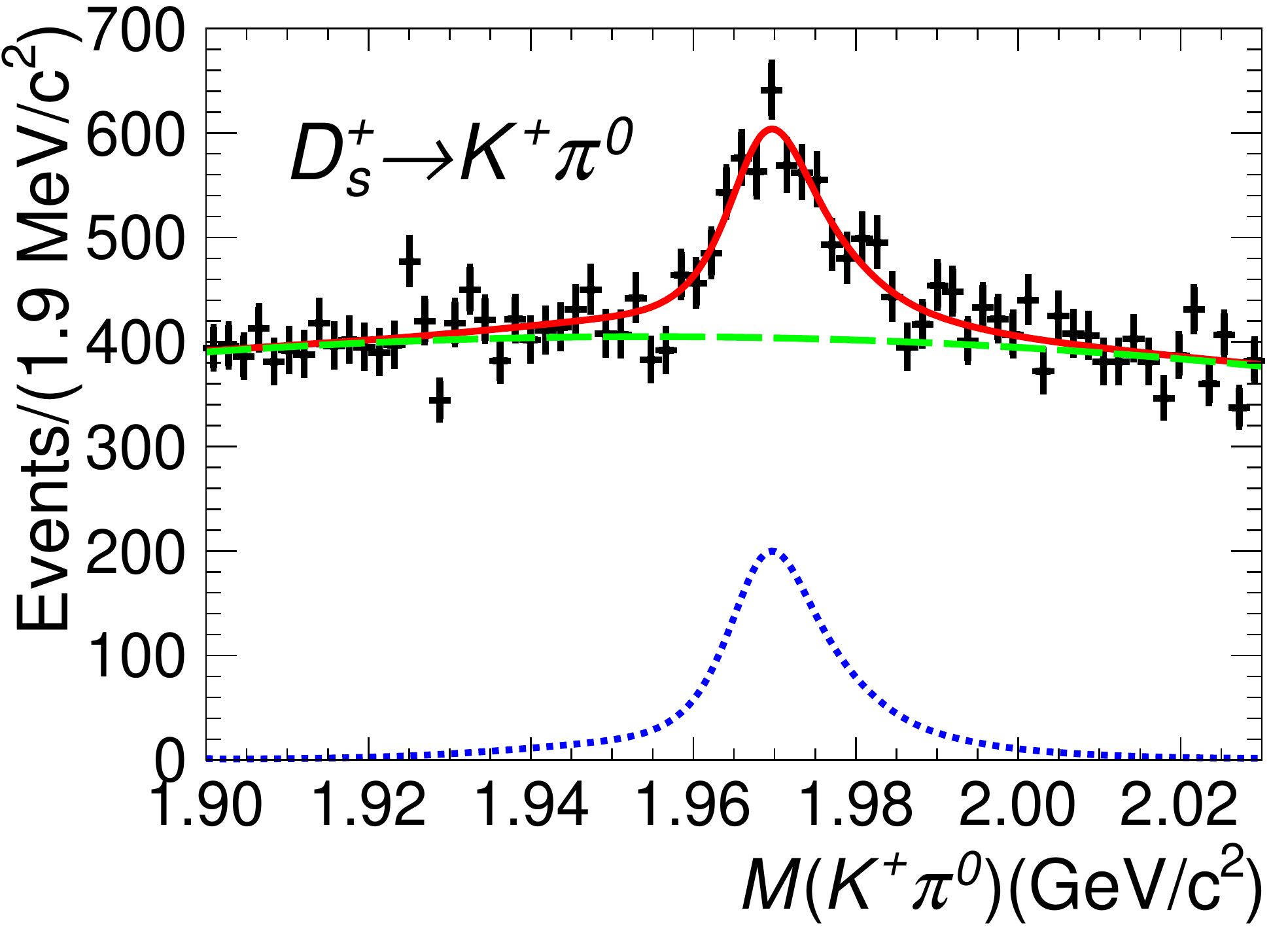}
		\end{minipage}
	}
	\subfigure{
		\begin{minipage}{0.4\columnwidth}
			\centering
			\includegraphics[width=2.5in]{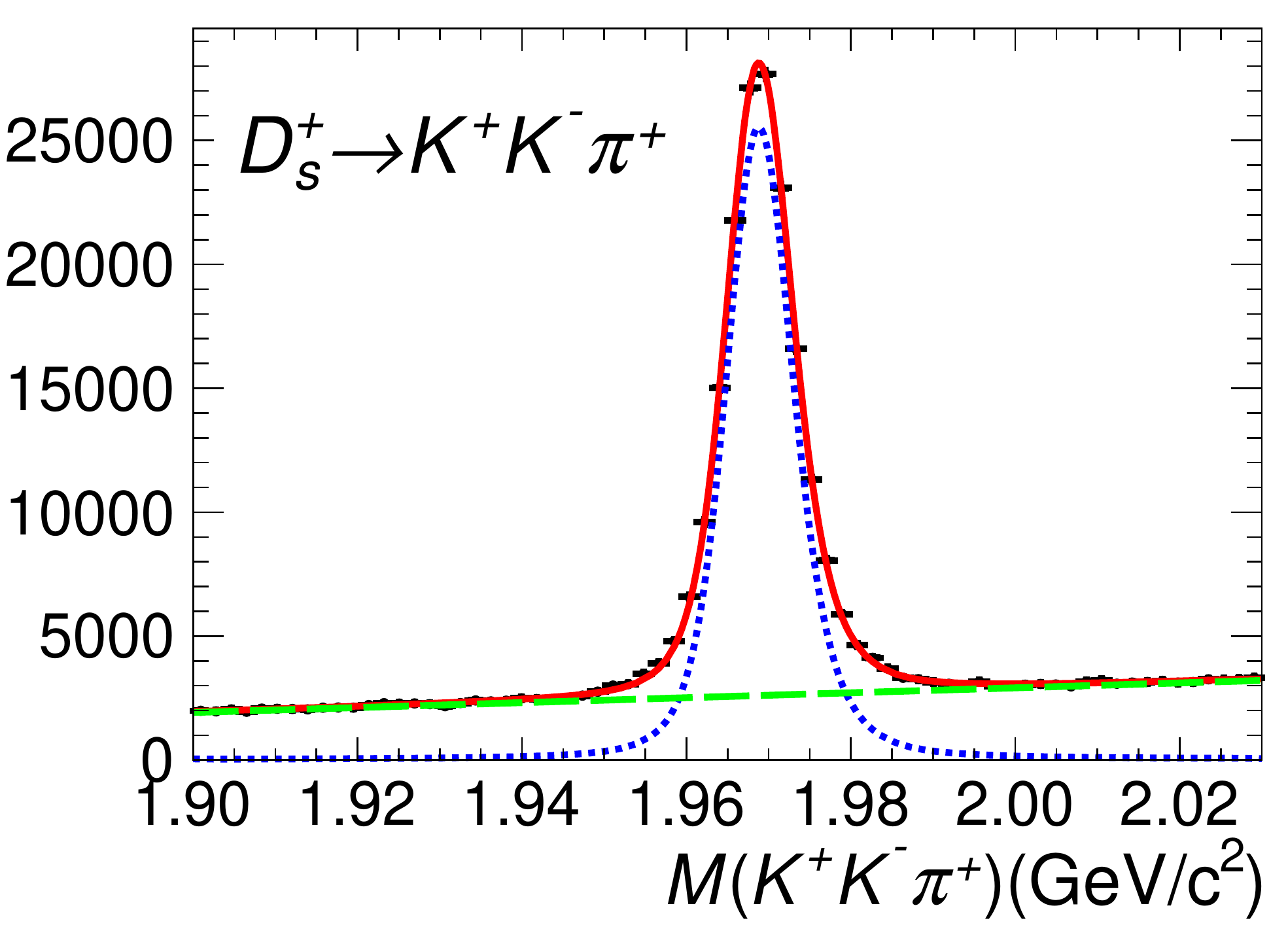}
		\end{minipage}
	}
	\caption{Fits to the invariant mass spectra of the signal candidates in data (shown as dots with error bars). The solid lines are the fit results, the dotted lines are the signal components, the long dashed lines are the non-peaking backgrounds and the dotted dashed lines are the peaking backgrounds.}
	\label{fig:data_yields}
\end{figure*}

To extract the signal yields for the signal $\dstopp$ decay modes and the normalization decay mode, unbinned extended maximum likelihood fits are performed on the $\mdsp$ distributions of the selected candidates in data. In each fit, the probability density function~(PDF) is parameterized as the sum of signal and background PDFs. The signal PDF is a template shape formed from the signal MC sample convolved with a Gaussian function to compensate the resolution difference between data and MC simulations. For the more common Cabibbo-favored (CF) decay modes $\dstokkpi$, $\dstokpipi$, $\dstopieta$, $\dstopietaprim$ and the SCS decay $\dstokpiz$, the Gaussian parameters are left free. 
For the low-yield SCS decays $\dstopipipi$, $\dstoketa$ and $\dstoketaprim$, the Gaussian parameters are fixed at the values obtained from the corresponding fits to the CF decay modes $\dstokpipi$, $\dstopieta$ and $\dstopietaprim$, respectively, since the kaons and pions have almost the same kinematics. According to the background study using inclusive MC samples, peaking backgrounds are present for the modes of $\dstopietaprim$, $\dstokpipi$ and $\dstopipipi$. The peaking backgrounds are modeled in the fit with the MC-determined shape and size. The fractions of the peaking background in the total event yields are estimated to be 2.0$\%$, 1.4$\%$ and 1.6$\%$ for $\dstopietaprim$, $\dstokpipi$ and $\dstopipipi$, respectively. The non-peaking background components are described with linear functions and second-order Chebychev functions for  the CF and SCS decay modes, respectively. The fits are presented in Fig.~\ref{fig:data_yields}, and the numerical results of the signal yields are listed in Table~\ref{tab:yield_efficiency_ratio}. The relative and absolute BFs, calculated with the average detection efficiencies obtained from the signal MC simulations, are summarized in Table~\ref{tab:yield_efficiency_ratio}. 
\begin{table}[h]
	\setlength{\abovecaptionskip}{0.cm}
	\setlength{\belowcaptionskip}{0.15cm}
	\begin{center}
		\caption{Summary of the signal yields, average detection efficiencies, relative BFs and absolute BFs of individual signal decay modes.  The first uncertainty is statistical, the second is systematic, and the third is external, from the BF of the normalization mode $\dstokkpi$~\cite{pdg2018}. The uncertainties on efficiencies are due to the limited MC event statistics.}
		\vspace{0.0cm}
		\footnotesize
		\newcommand{\tabincell}[2]{\begin{tabular}{@{}#1@{}}#2\end{tabular}}
		\begin{threeparttable}
			\begin{tabular}{c|c|c|c|c} \hline\hline
				Decay   &  $n^{i}$  &   $\overline{\varepsilon}^i$~(\%) & \emph{$R^{i}$}~(\%)   & $\mathcal{B}^i$~($10^{-3}$) \\
				\hline
				$\kaon^+\eta'$   &  $675\pm43$     &   $13.66\pm0.20$ &$4.91\pm0.31\pm0.31$    & $2.68\pm0.17\pm0.17\pm0.08$ \\
				$\eta'\pi^+$     &  $9912\pm113$   &   $14.19\pm0.04$ &$69.4\pm0.8\pm3.8$       & $37.8\pm0.4\pm2.1\pm1.2$\\
				$\kaon^+\eta$    &  $1841\pm114$   &   $26.21\pm0.17$ &$2.97\pm0.18\pm0.06$    & $1.62\pm0.10\pm0.03\pm0.05$\\
				$\eta\pi^+$      &  $19519\pm192$  &   $25.86\pm0.05$ &$31.94\pm0.33\pm0.49$    & $17.41\pm0.18\pm0.27\pm0.54$\\
				$\kaon^+\ks$     &  $35977\pm206$  &   $31.47\pm0.05$ &$27.55\pm0.18\pm0.50$    & $15.02\pm 0.10\pm0.27\pm 0.47$\\
				$\ks\pi^+$       &  $2724\pm83$    &   $32.27\pm0.16$ &$ 2.035\pm0.062\pm0.042$ & $1.109\pm 0.034\pm0.023\pm 0.035$\\
				$\kaon^+\piz$    &  $2275\pm149$   &   $27.96\pm0.18$ &$ 1.373\pm0.090\pm0.033$ & $0.748\pm 0.049\pm0.018\pm 0.023$\\
				$\kp\km\pi^+$    &  $160262\pm478$ &   $26.73\pm0.02$ &$100$                    & 54.5$\pm$1.7\\
				\hline\hline
			\end{tabular}
			\label{tab:yield_efficiency_ratio}
		\end{threeparttable}
	\end{center}
\end{table}

\section{SYSTEMATIC UNCERTAINTY}
The sources of systematic uncertainties considered in obtaining the relative BFs include the MC statistics, $\sigma$($e^+e^-\to\dsstpdsm$) lineshape, shapes of invariant mass distributions for signal and background, peaking background modeling, kinematic fit, $\deltaE$ and invariant mass requirements, reconstruction efficiency estimation and quoted BFs. Table~\ref{tab:sys_err} summarizes all of these systematic uncertainties. Some correlated uncertainties between the signal decay modes and the reference decay mode have been partially cancelled when extracting $\emph{$R^{i}$}$ in Table~\ref{tab:yield_efficiency_ratio}. 

\clearpage

\begin{table*}[tpb]
	\setlength{\abovecaptionskip}{0.0cm}
	\setlength{\belowcaptionskip}{-1.6cm}
	\caption{Summary of the systematic uncertainties~(in unit of $\%$) for the measurements of relative BFs . The total values are calculated by summing up all contributions in quadrature.}
	\begin{center}
		\newcommand{\tabincell}[2]{\begin{tabular}{@{}#1@{}}#2\end{tabular}} 
		\begin{threeparttable}
			\begin{tabular}{c|c c c c c c c}
				\hline \hline
				Source & $\kaon^+\eta'$ & $\eta'\pi^+$ &$\kaon^+\eta$ & $\eta\pi^+$ & $\kaon^+\ks$ & $\ks\pi^+$ & $\kaon^+\pi^0$  \\
				\hline
				MC statistics      & 0.7 & 0.1 & 0.3 & 0.1 & 0.1 & 0.2 & 0.3 \\
				Lineshape          & 1.0 & 0.5 & 1.1 & 0.9 & 0.1 & 1.0 & 1.8 \\
				Signal shape       & 1.0 & 1.0 & 0.7 & 0.7 & 0.3 & 0.3 & 0.3 \\
				Background shape   & 0.0 & 0.3 & 1.0 & 0.2 & 0.0 & 0.8 & 1.4 \\
				Peaking background &  -  & 0.8 &  -  &  -  & 0.0 & 0.1 &  -  \\
				Kinematic fit      & 0.6 & 0.6 & 0.6 & 0.6 & 0.0 & 0.0 & 0.6 \\
				$\deltaE$ and invariant masses 
				& 2.2 & 1.8 & 0.4 & 0.4 & 1.1 & 1.0 & 0.4 \\
				\tabincell{c}{Reconstruction efficiency} 
				& 5.4 & 4.6 & 0.2 & 0.5 & 1.4 & 1.2 & 0.0 \\
				Quoted BFs         & 1.7 & 1.7 & 0.5 & 0.5 & 0.1 & 0.1 & 0.0 \\	
				\hline
				Total & 6.3 & 5.4 & 1.9 & 1.5 & 1.8 & 2.1 & 2.4 \\
				\hline\hline
			\end{tabular}
			\label{tab:sys_err}
			
		\end{threeparttable}
		
	\end{center}
\end{table*}

\begin{itemize}
	\item \emph{MC Statistics.} Average detection efficiencies are evaluated using MC simulated samples. The uncertainties due to the limited sample sizes, obtained by propagating the statistical uncertainties of the individual efficiencies at different energy points according to Eq.~\eqref{eq:eff_weight}, are assigned as systematic uncertainties.
	
	\item \emph{$\sigma${\rm (}$e^+e^-\to\dsstpdsm${\rm )} lineshape.}  Signal PDFs and detection efficiencies have slight dependencies on the input lineshape of $\sigma$($e^+e^-\to\dsstpdsm$). To evaluate this uncertainty, different lineshapes are used to estimate the detection efficiencies and data yields. The resulting changes in BFs are taken as systematic uncertainties.
	
	\item \emph{Signal shape.} The uncertainties related to the signal shapes are studied using the decays $\dstokpiz$, $\dstopietaprim$, $\dstopieta$ and $\dstokpipi$. In the nominal analysis, signal shape in the $\mdsp$ distribution of the signal candidates is modeled by the signal PDF convolved with a Gaussian function.  Double-Gaussian functions are used instead as convolution functions, and the resultant changes of BFs are taken as systematic uncertainties. For the low-yield SCS decays $\dstopipipi$, $\dstoketa$ and $\dstoketaprim$ the uncertainties of the corresponding CF modes are used.
	
	\item \emph{Background shape.} In the nominal analysis, the background shapes are described by first-order polynomial functions for the decays $\dstopietaprim$, $\dstopieta$, $\dstokpipi$ and $\dstokkpi$ and second-order polynomials for the decays $\dstoketaprim$, $\dstoketa$, $\dstopipipi$ and $\dstokpiz$. To estimate the uncertainties from the background shapes, higher-order polynomials are considered as alternatives: second-order and third-order, respectively. The resulting changes of the BFs are taken as systematic uncertainties.
	
	\item \emph{Peaking background.} The contributions to the peaking backgrounds of $\dstokpipi$,  $\dstopipipi$ and $\dstopietaprim$ are from the decays of $D^+\to\ks\pi^+$ (due to $\kaon^{+}$ and $\pi^{+}$ misidentification),  $D_s^+\to\pi^+\pi^+\pi^-$ and $D_s^+\to a_1(1260)^+\eta~(a_1(1260)^+\to\rho^0\pip,\rho^0\to\pip\pim)$~\cite{Ablikim:preliminary}, respectively. Their shapes and sizes are fixed according to MC simulations in the fit. The input BFs of these background processes are varied by their uncertainties and the changes in results are taken as systematic uncertainties.  
	
	\item \emph{Kinematic fit.} High-yield CF decays of $\dstokpipi$ and $\dstopieta$ are used to study the uncertainty due to the kinematic fit. We perform the analysis without applying the kinematic fit.  The differences from the nominal results  are  taken as systematic uncertainties. For the $\dstopipipi$ mode the uncertainty from $\dstokpipi$ is taken while the uncertainty from $\dstopieta$ is assigned to the decays with photons in the final states.
	
	\item \emph{$\deltaE$ and invariant mass requirements.} To estimate potential bias on efficiency estimations by restricting the kinematics in the selected regions, the distributions of the kinematic variables in MC simulations are smeared with Gaussian functions. The parameters of the functions are obtained by fitting the smeared MC distributions to the corresponding distributions in data. The variables $\deltaE$, ${\it M}$($\pip\pim$), ${\it M}$($\gamma\gamma$), ${\it M}$($\pip\pim\eta$), ${\it M_{\rm rec}}(D_{s}^{+})$ and ${\it M}(D_{s}^{+}\gamma)$ are studied. Updated efficiencies based on the Gaussian-smeared MC simulations are obtained and the relative changes from the nominal efficiencies are assigned as the systematic uncertainties.
	
	\item \emph{Reconstruction efficiency.} We consider the efficiencies of tracking and PID~($K^{\pm}$, $\pi^{\pm}$) and the efficiencies of intermediate particles~($\pi^{0}$, $\eta$, $\ks$) reconstructions, which are studied based on a series of control samples.
	The $\kaon^{\pm}$ and $\pi^{\pm}$ tracking and PID efficiencies are studied using control samples of $e^+e^-\to\kaon^+\kaon^-\pi^+\pi^-$, $\kaon^+\kaon^-\kaon^+\kaon^-$, $\kaon^+\kaon^-\pi^+\pi^-\piz$, $\pi^+\pi^-\pi^+\pi^-$ and $\pi^+\pi^-\pi^+\pi^-\piz$ events~\cite{Ablikim:2019whl}.
	A partial cancellation of the tracking and PID uncertainties in the ratio of the signal modes and the normalization mode is taken into account.
	The $\piz$ and $\eta$ reconstruction efficiencies are evaluated using the double-tag $D\bar{D}$ hadronic decays $D^0\to\kaon^-\pi^+$, $\kaon^-\pi^+\pi^+\pi^-$ versus $\bar{D}^0\to\kaon^+\pi^-\piz$, $\ks\piz$~\cite{Ablikim:2016sqt, Ablikim:2016xny} and approximating the $\eta$ behavior as similar to the $\pi^0$. The $\ks$ reconstruction efficiency is studied with samples of $J/\psi\to\kaon^{*}(892)^{\pm}\kaon^{\mp}$, $\kaon^{*}(892)^{\pm}\to\ks\pi^{\pm}$ and $J/\psi\to\phi\ks\kaon^{\mp}\pi^{\pm}$~\cite{Ablikim:2015qgt}.
	To account for the different kinematics of the various signal modes, the nominal detection efficiencies are scaled based on event-by-event corrections according to the momentum-dependent efficiency differences between MC simulations and data. The appropriately averaged scaling factors are assigned as the corresponding systematic uncertainties, as given in Table~\ref{tab:sys_err}. Here, the $\dstoketaprim$ and $\dstopietaprim$ decays suffer from large reconstruction uncertainties due to the low-momentum charged pions and $\eta$ from $\eta'$ decay.
	
	\item \emph{Quoted BFs.} The nominal BFs of $\ks\to\pi^{+}\pi^{-}$, $\pi^{0}\to\gamma\gamma$, $\eta\to\gamma\gamma$ and $\eta'\to\eta\pi^{+}\pi^{-}$ are used and their corresponding uncertainties~\cite{pdg2018} are propagated as systematic uncertainties.  
\end{itemize}

\begin{table}[h]
	\caption{ Results of the obtained relative BFs (in unit of $\%$). The first uncertainty is statistical,  and the second is systematic.}
	\label{tab:other_branching_ratios}
	\begin{center}
		\begin{threeparttable}
			\begin{tabular}{c| c c }
				\hline\hline
				Relative BFs & This work & PDG~\cite{pdg2018} \\
				\hline
				$\mathcal{B}(\kaon^+\eta')$/$\mathcal{B}(\eta'\pi^+)$    & $7.07\pm0.46\pm0.11$  &  $4.2\pm1.3$    \\
				$\mathcal{B}(\kaon^+\eta)$/$\mathcal{B}(\eta\pi^+)$      & $9.31\pm0.58\pm0.10$  &  $8.9\pm1.6$    \\
				$\mathcal{B}(\ks\pi^+)$/$\mathcal{B}(\kaon^+\ks)$        & $7.38\pm0.23\pm0.09$  &  $8.12\pm0.28$  \\
				$\mathcal{B}(\kaon^+\eta)$/$\mathcal{B}(\kaon^+\eta')$   & $60.6\pm5.4\pm3.6$    &               --\\
				$\mathcal{B}(\eta\pi^+)$/$\mathcal{B}(\eta'\pi^+)$       & $46.0\pm0.7\pm2.1$    &               --\\
				\hline\hline
			\end{tabular}
		\end{threeparttable}
	\end{center}
\end{table}

\section{SUMMARY AND DISCUSSION}
The BFs for $\dstoketaprim$, $\dstopietaprim$, $\dstoketa$, $\dstopieta$, $\dstokpipi$, $\dstopipipi$ and $\dstokpiz$ are measured using $\ee$ collision data collected at $\sqrt{s}=4.178\sim4.226~\gev$ in the BESIII experiment. 
The results obtained in this work are listed in Table ~\ref{tab:yield_efficiency_ratio}
and can be compared with the results from PDG~\cite{pdg2018} as well as with theoretical predictions 
\cite{Hai-Yang Cheng2010, Fu-Sheng Yu2011, Hsiang-nan Li2012, Di Wang2017} (Table~\ref{tab:results_BFs_theo}).
Our results are consistent with the PDG values, while the precision is three to five times better than that of previous results. In addition, our results in general agree with the available theoretical calculations~\cite{Cheng:2019ggx,Hai-Yang Cheng2010,Fu-Sheng Yu2011,Hsiang-nan Li2012,Di Wang2017} within about 3$\sigma$. However, the discrepancies from our measurements are significant for the model calculations in Ref.~\cite{Hai-Yang Cheng2010} for the modes $\dstoketaprim$ and $\dstoketa$, and from the model calculations in Ref.~\cite{Hsiang-nan Li2012} for the mode $\dstoketa$. Investigating these discrepancies should aid in further developing these QCD-derived models in charm physics.

The ratios of the BFs, $\mathcal{B}(\kaon^+\eta')$/$\mathcal{B}(\eta'\pi^+)$, $\mathcal{B}(\kaon^+\eta)$/$\mathcal{B}(\eta\pi^+)$, $\mathcal{B}(\ks\pi^+)$/ $\mathcal{B}(\kaon^+\ks)$, $\mathcal{B}(\kaon^+\eta)$/$\mathcal{B}(\kaon^+\eta')$, and $\mathcal{B}(\eta\pi^+)$/$\mathcal{B}(\eta'\pi^+)$, are also determined, as listed in Table~\ref{tab:other_branching_ratios}. The partial cancellations of the systematic uncertainties from $\sigma$($e^+e^-\to\dsstpdsm$) lineshape, signal shape, background shape, peaking background, kinematic fit, $\deltaE$ and invariant mass requirements, and reconstruction efficiency between the pairs of decay modes are considered. Our results of $\mathcal{B}(\kaon^+\eta')$/$\mathcal{B}(\eta'\pi^+)$, $\mathcal{B}(\kaon^+\eta)$/$\mathcal{B}(\eta\pi^+)$, $\mathcal{B}(\ks\pi^+)$/ $\mathcal{B}(\kaon^+\ks)$ are consistent with the PDG values within about 2$\sigma$, but the precisions are improved. Our results are also in general accord with the theoretical calculations~\cite{Cheng:2019ggx,Hai-Yang Cheng2010,Fu-Sheng Yu2011,Hsiang-nan Li2012,Di Wang2017} within about 3$\sigma$. However, our measurements are in disagreement with the model calculations in Refs.~\cite{Hai-Yang Cheng2010, Fu-Sheng Yu2011} for the ratio $\mathcal{B}(\kaon^+\eta')$/$\mathcal{B}(\eta'\pi^+)$ and with those in Refs.~\cite{Hai-Yang Cheng2010, Hsiang-nan Li2012} for the ratio $\mathcal{B}(\kaon^+\eta)$/$\mathcal{B}(\eta\pi^+)$. 
\noindent The theoretical uncertainties on these ratios are expected to be reduced as well, offering more meaningful comparisons between experimental measurements and theoretical calculations.

\acknowledgments

The BESIII collaboration thanks the staff of BEPCII and the IHEP computing center for their strong support. This work is supported in part by National Natural Science Foundation of China (NSFC) under Contracts Nos. 11625523, 11635010, 11675275, 11735014, 11775027, 11822506, 11835012, 11935015, 11935016, 11935018, 11975021, 11961141012; National Key Basic Research Program of China under Contract No. 2015CB856700; the Chinese Academy of Sciences (CAS) Large-Scale Scientific Facility Program; CAS Key Research Program of Frontier Sciences under Contracts Nos. QYZDJ-SSW-SLH003, QYZDJ-SSW-SLH040; Joint Large-Scale Scientific Facility Funds of the NSFC and CAS under Contracts Nos. U1932101, U1832207, U1732263; 100 Talents Program of CAS; National 1000 Talents Program of China; INPAC and Shanghai Key Laboratory for Particle Physics and Cosmology;
The University of Chinese Academy of Sciences; 
The Beijing municipal government under Contract No. CIT$\&$TCD201704047;
ERC under Contract No. 758462; German Research Foundation DFG under Contracts Nos. Collaborative Research Center CRC 1044, FOR 2359; Istituto Nazionale di Fisica Nucleare, Italy; Ministry of Development of Turkey under Contract No. DPT2006K-120470; National Science and Technology fund; STFC (United Kingdom); The Knut and Alice Wallenberg Foundation (Sweden) under Contract No. 2016.0157; The Royal Society, UK under Contracts Nos. DH140054, DH160214; The Swedish Research Council; U. S. Department of Energy under Contracts Nos. DE-FG02-05ER41374, DE-SC-0012069.

\end{document}